\documentclass[aps,prd,preprint,eqsecnum,nofootinbib,showkeys,%
showpacs,groupedaddress,a4paper,byrevtex]{revtex4-1}
\usepackage{bm}
\usepackage{amsmath}
\usepackage{amsfonts,amssymb}
\begin{document}
\newcommand{\W}{\boldsymbol{\mathfrak{W}}}
\newcommand{\rmi}{\text{i}}
\newcommand{\tm}{T^{\,\text{M}}_{\alpha \beta}}
\newcommand{\ta}{T^{\text{A}}_{\alpha \beta}}
\newcommand{\tsym}{T^{\text{\,sym}}_{\alpha \beta}}
\newcommand{\g}{\ensuremath{\overline\Gamma}}
\newcommand{\wt}{\widetilde{\Omega}}
\title{Ponderomotive forces in electrodynamics of moving media: \\ The Minkowski and Abraham approaches}
\author{V.V.~Nesterenko}
\email[E-mail:~]{nestr@theor.jinr.ru}
\author{A.V.~Nesterenko}
\affiliation{Bogoliubov Laboratory of Theoretical Physics,
Joint Institute for Nuclear Research, \\
141980 Dubna, Russia}
\date{\today}
\begin{abstract}
In the general setting of the problem, the explicit compact
formulae are derived for the ponderomotive forces in the
macroscopic electrodynamics of moving media in the
Minkowski and Abraham approaches. Taking account of the
Minkowski constitutive relations and making use  of a
special representation for the Abraham energy-momentum
tensor enable one to obtain a compact expression for the
Abraham force in the case of arbitrary dependence of the
medium velocity on spatial coordinates and the time and for
nonstationary external electromagnetic field. We term the
difference between the ponderomotive forces in the Abraham
and Minkowski approaches as the Abraham force not only
under consideration of media at rest but also in the case
of moving media. The Lorentz force is found which is
exerted by external electromagnetic field on the conduction
current in a medium, the covariant Ohm law and the
constitutive Minkowski relations being taken into account.
The physical argumentation is traced for definition of the
4-vector of the ponderomotive force as the 4-divergence of
the energy-momentum tensor of electromagnetic field in a
medium.
\end{abstract}
\maketitle
\tableofcontents
\section{Introduction}
\label{Intr} In macroscopic electrodynamics, there is no
generally accepted definition of the ponderomotive forces.
Usually  this problem is bound up with the definition of
the energy-momentum tensor of electromagnetic field in a
medium, with presupposing the ponderomotive force being
equal to the four-dimensional divergence of this tensor and
with taking into account the Maxwell equations. Up to now
the use of the theoretical argumentation did not allow one
to define uniquely the energy-momentum tensor in medium.
Therefore when analyzing the ponderomotive force, the
working formulae are useful which are derived by making use
of the  different versions of this tensor. It is this
problem that is considered in the present paper. We derive
compact 3-dimensional vector formulae for ponderomotive
forces in the case of moving media by making use of the
electromagnetic energy-momentum tensor in the Minkowski
form and in the Abraham form. These tensors proposed more
than one century ago are the most popular and are often
discussed~\cite{Brevik-1,Brevik-3,NN-A}. In the
framework of other approaches they are treated as
distinctive ``fulcra'' (see, for example, Ref.\
\cite{Nelson}).

We do not touch  the problem of determining the `correct'
energy-momentum tensor in macroscopic electrodynamics and
the Abraham-Minkowski controversy. There is a large body of
the literature on this subject, see, for example, the
reviews \cite{Brevik-1,Brevik-3,Groot,MR,Obukhov} and
references therein.

  In macroscopic electrodynamics, the important role is
played by the constitutive relations between field vectors.
Only due to  these relations the formal scheme of this
theory acquires the physical content~\cite[Sec.\
39]{Pauli}. We shall use the constitutive relations in the
form derived by Minkowski. The values of the vectors
$\mathbf{E,H,D,B}$ and the velocity of the medium
$\mathbf{v}, v<c$, which satisfy the constitutive relations
will be refereed to as the physical configuration space
$\g$ in contrast to the whole unrestricted configuration
space $\Gamma$. In practical calculations aimed at the
obtaining of the final physical results, one assumes that the constitutive
relations will be eventually used. Therefore in
deriving the formulae for the ponderomotive forces it makes
sense to take into account, aside from the Maxwell
equations, the constitutive relations too. Minkowski
\cite{M-zwei} and Abraham \cite{AP1909,AP1910,Abraham1920}
proceeded just in this way. This simplifies essentially the
ultimate formulae for moving media, however the realization
of this task turns out to be rather complicated.

In the general case, the ponderomotive forces include those
of two types, the Lorentz force experienced by free charges
and currents, and the forces exerted by electromagnetic
field on a medium. When deriving the Lorentz force  from
the divergence  of the energy-momentum tensor, the Maxwell
equations are used. The remaining part of the divergence is
transformed with allowance for the constitutive equations
resulting in the formulae for the ponderomotive forces
defined on $\g$.

As known, in the case of a medium at rest the Abraham
energy-momentum tensor gives an additional, as  compared
with the Minkowski approach, term for the ponderomotive
force, the  so-called Abraham force. This force is
different from zero only for the external field dependent
on time. In the case of moving media the difference between
the Abraham approach and Minkowski approach  turns out to
be more substantial because the Abraham energy-momentum
tensor, in contrast to the Minkowski tensor, depends
explicitly on the medium velocity. For practical use, it
makes sense to derive such formulae for ponderomotive
forces in which this distinction is shown clearly and the formulae
themselves are simple as much as possible. Further we shall
call the difference between the ponderomotive forces in the
Abraham and Minkowski approaches as the Abraham force both in
the case of media at rest and in the case of moving media.

The making use of a special representation for the Abraham
energy-momentum tensor on $\g$, which is the most close, in
structure, to the Minkowski tensor, will enable us to
obtain a compact formula for the Abraham force in the case
of moving media with an arbitrary dependence of the medium
velocity on the coordinates and time and for the time
dependent external electromagnetic field.

 Before beginning the calculation of the ponderomotive
forces, we consider the physical reasoning which is adduced
in favour of defining these forces through the divergence
of the energy-momentum tensor of electromagnetic field in
medium. In particular, it will be shown that this
argumentation is insufficient for defining this tensor
uniquely.

The layout of the paper is as follows. In Sec.\ \ref{edm},
the Minkowski electrodynamics of moving media is briefly
stated. In Sec.\ \ref{ftei}, the physical reasoning is
retraced which leads to the definition  of the 4-vector of
the ponderomotive force in terms of the divergence of the
electromagnetic energy-momentum tensor in medium. In Sec.\
\ref{Min}, the formulae for the ponderomotive forces are
derived on $\g$ in the framework of the Minkowski approach.
The general setting of the problem is considered, namely,
nonhomogeneous moving medium with arbitrary dependence of
its velocity on coordinates and time and the nonstationary
external electromagnetic field. In addition, the
macroscopic Lorentz force is found here which is exerted by
external electromagnetic field on the conduction current in
a medium, the covariant Ohm law and the constitutive
Minkowski relations being taken into account. In Sec.\
\ref{Ab-dif} the different representations of the Abraham
energy-momentum tensor on $\g$ are discussed. In Sec.\
\ref{Ab-f} the formulae are derived for the ponderomotive
forces in the Abraham approach. For that a special
representation of the Abraham tensor is employed which is
the most close, according to its structure, to the
Minkowski energy-momentum tensor on $\g$. In Conclusion
(Sec.\ \ref{Concl}) the obtained results are briefly
summarized. In Appendix \ref{A} in the framework of the
electrodynamics in vacuum, the interrelation is traced
between the ponderomotive forces (in the present case
between the Lorentz force) and the energy-momentum tensor
of electromagnetic field in vacuum. In  Appendix
\ref{B}, the explicit formula is derived for the vector $\W
$ the making use of which enables one to represent, in a
compact  form, the ultimate formulae for the ponderomotive
forces on~$\g$.

We use the following notations. The coordinates $x_\alpha$
in the Minkowski space have a pure imaginary time
component: $x_\alpha=(x_1,x_2,x_3,x_4 =\rmi ct)=
(\mathbf{x}, \rmi c t)$. The Greek indices take values
$1,2,3,4$, whereas the Latin indices assume the values
$x,y,z$ or $1,2,3$; the summation over the repeated indexes
is understood in respective range. The unrationalized
Gaussian units are used for the electromagnetic field and
the notations generally accepted in macroscopic
electrodynamics \cite{LL8} are adopted.

\section{Minkowski macroscopic electrodynamics of moving media}
\label{edm} We shall consider the electromagnetic field in
a material media in the framework of the Minkowski
phenomenological electrodynamics \cite[Secs.\
33--35]{Pauli}. Here we give the basic equations defining
this theory. The electromagnetic field in a medium is
described by the macroscopic Maxwell equations
\begin{gather}
\text{rot}\,\mathbf{E}+\frac{1}{c}\,\frac{\partial\,\mathbf{B}}{\partial t}=0,
\quad
\text{div}\,\mathbf{B}=0\,{,} \label{m-1} \\
\text{rot}\,\mathbf{H}-\frac{1}{c}\,\frac{\partial\,\mathbf{D}}{\partial t}= \frac{4\pi}{c}\,\mathbf{j}\,{,}
\quad
\text{div}\,\mathbf{D}=4\pi \rho\,{.} \label{m-2}
\end{gather}
In a covariant form these equations read
\begin{gather}
 \frac{\partial F_{\alpha \beta}}{\partial
x_\gamma}+\frac{\partial F_{ \beta \gamma}}{\partial
x_\alpha}+\frac{\partial F_{\gamma \alpha }}{\partial
x_\beta}=0\, {,} \label{m-5}\\
\frac{\partial H_{\alpha \beta}}{\partial x_\beta}=\frac{4\pi}{c}j_\alpha\,{.}
\label{m-6}
\end{gather}
Here $F_{\alpha \beta}$ and $H_{\alpha \beta}$ are two
antisymmetric tensors introduced by Minkowski
\cite[Sec.~33]{Pauli},
\cite[\S~76]{LL8},
\begin{gather}
F_{ij}=\varepsilon_{ijk}B_k,\quad F_{4j}=-F_{j4}=\rmi E_{j}\,{;}
\label{m-7} \\
H_{ij}=\varepsilon_{ijk}H_k,\quad
H_{4j}=-H_{j4}=\rmi D_{j}\,{.} \label{m-8}
\end{gather}
It is convenient to represent Eqs.\ \eqref{m-7} and
\eqref{m-8} in a short notation, explicitly indicating the
3-dimensional vectors, which define the components of the
antisymmetric tensors~\cite{M-zwei,Kafka,LL2}:
\begin{equation}
\label{m-8a} F_{\alpha \beta}(\mathbf{B};-\rmi
\mathbf{E}),\quad H_{\alpha \beta}(\mathbf{H};-\rmi
\mathbf{D})\,{.}
\end{equation}

In the general case, the current density in nonhomogeneous
Maxwell equations \eqref{m-2}, \eqref{m-6}
\begin{equation}
\label{m-9}
j_\alpha=(\mathbf{j},\rmi c\rho)
\end{equation}
is the sum of two terms, the density of the conduction
current and the density of convection current (see
Eq.\ \eqref{cond} and further). The total current density
$j_\alpha$ satisfies the continuity equation
\begin{equation}
\label{m-10} \text{div\,} \mathbf{j}+\frac{\partial
\rho}{\partial t}\equiv\frac{\partial j_\alpha}{\partial
x_\alpha}=0\,{,}
\end{equation}
which entails the conservation of the charge. Equation
\eqref{m-10} is the consistency condition  of the Maxwell
equations \eqref{m-2}, \eqref{m-6}.

The Maxwell equations \eqref{m-1}--\eqref{m-6} and
continuity equation \eqref{m-10} constitute a formal
mathematical scheme, which is not filled with the physical
content yet. Indeed, the number of these equations in the
material medium and in vacuum is the same, whereas the
number of the field functions in a medium is two times as
large as  compared with vacuum. Further, the equations in
question do not contain the material characteristics of
medium. These problems are removed by introducing the
constitutive relations which play an important role in
macroscopic electrodynamics.

We shall consider a linear isotropic medium without
temporal and spatial dispersion with the constitutive relations
\begin{equation}
\label{e0}
\mathbf{D}=\varepsilon \mathbf{E},\quad  \mathbf{B}=\mu\mathbf{H}
\end{equation}
(medium at rest). In order to generalize \eqref{e0} to
moving medium, Minkowski introduced the Lorentz vectors
$E_\alpha,D_\alpha$ which in the co-moving reference frame, $K'$,
where the medium is at rest ($\mathbf{v}'=0$), have the components
\begin{equation}
\label{rest}
E'_\alpha=(\mathbf{E}',0), \quad
D'_\alpha=(\mathbf{D}',0), \quad
H'_\alpha=(\mathbf{H}',0), \quad
B'_\alpha=(\mathbf{B}',0)\, {.}
\end{equation}
The prime implies that the respective quantity is
considered in the co-moving frame. Obviously  conditions
\eqref{rest} determine the vectors $E_\alpha,D_\alpha$ and
$H_\alpha,B_\alpha$ uniquely. Minkowski  \cite{M-zwei}
constructed for these vectors the explicit formulae (see
also~\cite{Kafka,Laue,Groot})
\begin{gather}
E_\alpha= F_{\alpha \beta}u_\beta=\gamma\{
\mathbf{E}+[\mathbf{q}\mathbf{B}],\;\rmi(\mathbf{q}\mathbf{E})
\}\,{,}\quad  D_\alpha = H_{\alpha \beta}u_{\beta}=\gamma\{
\mathbf{D}+[\mathbf{q}\mathbf{H}],\;\rmi(\mathbf{q}\mathbf{D})
\}\,{,} \label{Kf-1}\\
H_\alpha=-\rmi H^*_{\alpha \beta}u_{\beta}=\gamma\{
\mathbf{H}-[\mathbf{q}\mathbf{D}],\;\rmi(\mathbf{q}\mathbf{H})
\}\,{,}\quad B_\alpha =-\rmi F^*_{\alpha \beta}u_{\beta}=\gamma\{
\mathbf{B}-[\mathbf{q}\mathbf{E}],\;\rmi(\mathbf{q}\mathbf{B})
\}{.}\label{Kf-2}
\end{gather}
Here  $u_\alpha$ is the 4-velocity of the medium
\begin{equation}
\label{eu} u_\alpha=\gamma\{\mathbf {q},
\text{i}\},\quad  \mathbf {q}= \mathbf{v}/c,\quad
\gamma^{-1}=\sqrt{1-\mathbf{q}^2}, \quad u_\alpha
u_\alpha=-1, \quad \alpha =1,2,3,4\,{,}
\end{equation}
and  $F^*_{\alpha \beta}$ and $H^*_{\alpha \beta}$ are the
tensors which are dual to the tensors $F_{\alpha \beta}$
and $H_{\alpha \beta}$ respectively:
\begin{equation}
\label{dual} F^*_{\alpha
\beta}=\frac{1}{2}\,\varepsilon_{\alpha \beta \gamma
\delta}F_{\gamma \delta},\quad  H^*_{\alpha
\beta}=\frac{1}{2}\,\varepsilon_{\alpha \beta \gamma
\delta}H_{\gamma \delta},\quad \varepsilon _{1\,2\,3\,4}=+1\,{.}
\end{equation}
By making use of notation  \eqref{m-8a}, we can write
\begin{equation}
\label{dual-1}
-\rmi F^*_{\alpha
\beta}(-\mathbf{E};-\rmi \mathbf{B});\quad -\rmi H^*_{\alpha
\beta}(-\mathbf{D};-\rmi \mathbf{H})\,{.}
\end{equation}
From  definitions \eqref{Kf-1} and \eqref{Kf-2} it follows
that the vectors $E_\alpha,\,D_\alpha$ and
$H_\alpha,\,B_\alpha$ are orthogonal to the medium velocity $u_\alpha$
\begin{equation}
\label{orth}
E_\alpha u_\alpha=0,\quad
D_\alpha u_\alpha=0,\quad
H_\alpha u_\alpha=0,\quad
B_\alpha u_\alpha=0
\end{equation}
and   at $\mathbf{q}=0$ satisfy condition \eqref{rest}. In
the covariant form the constitutive relations \eqref{e0} are written as
\begin{eqnarray}
D_\alpha&=&\varepsilon E_{\alpha}\, {,} \label{Min-1}\\
B_{\alpha}&=&\mu H_{\alpha}
\label{Min-2}
\end{eqnarray}
or in the 3-dimensional vector notation \cite[Sec.\
33]{Pauli}, \cite[\S~76]{LL8}
\begin{gather}
\mathbf{D}+[\mathbf{q} \mathbf{H}]=\varepsilon\left (
\mathbf{E}+[\mathbf{q} \mathbf{B}]
\right ),\label{Min-3}\\
\mathbf{B}-[\mathbf{q}\mathbf{E}]=\mu\left (
\mathbf{H}-[\mathbf{q}\mathbf{D}]
\right ){.}\label{Min-4}
\end{gather}
As noted in  Introduction, the values of the field
variables $\mathbf{E,\; H,\;D,\; B}$ and the velocity of a
medium\footnote{We do not consider the case when the medium
velocity $\mathbf{v}$ is equal to the phase velocity of
light in the given medium, $v^2=c^2/(\varepsilon \mu)$. For
this value of $v$ the equations \eqref{Min-3} and
\eqref{Min-4} cannot be resolved with respect to $\mathbf{D}$ and $\mathbf{B}$
(see  \cite[Sec.\ 33]{Pauli}).} $\mathbf{v}, \; v<c$, which satisfy the constitutive
Minkowski relations (\ref{Min-1})--(\ref{Min-4}), will be
referred to as {\it the physical configuration space} $\g$
\cite{NN-A}.

The material relation for the current density \eqref{m-9}
is derived in a similar way \cite{M-zwei}. As noted above,
in the general case the current density $j_\mu$ is the sum
of the density of  conduction current and  the density of
convection current,
\begin{equation}
j_\mu=j^{\text{cond}}_\mu+j^{\text{conv}}_\mu{.}
\label{cond}
\end{equation}
In the co-moving frame $K'$, where the medium is at rest
 $(\mathbf{v}=0)$, these densities, by definition, have the
components~\cite[\S~60]{Becker}
\begin{align}
(j^{\text{\,cond}}_\mu)'=(\mathbf{j'}_{\text{cond}},0)\,{,}
\label{cond-1} \\
(j^{\text{\,conv}}_\mu)' =(\bm{0},\rmi c\rho'_{\text{conv}})\,{.}
\label{cond-2}
\end{align}
For simplicity, we introduce the notation~\cite{Becker,Moller}
\begin{equation}
\label{cond-3}
\mathbf{j}^{\,\prime}_{\text{cond}}=\mathbf{j}^{\,0}, \quad\rho'_{\text{conv}}=\rho^0{.}
\end{equation}
In the rest frame, we have for the total current \eqref{cond}
\begin{equation}
\label{cond-4}
{j}^{\,\prime}_\mu=(j_\mu^{\text{cond}})'+(j_\mu^{\text{conv}})'
=(\mathbf{j}^{\,0},\rmi c \rho^{\,0})\,{.}
\end{equation}

By making use of the Lorentz transformations the total
current \eqref{cond}, in an arbitrary inertial reference
frame $K$, can be found through   its components
\eqref{cond-4} in the rest frame. We accomplish this
transformation separately for the conduction current
\eqref{cond-1} and for the convection current \eqref{cond-2}.
In the first case we have
\begin{equation}
\label{cond-5}
{j}^{\text{\,cond}}_\mu=(\mathbf{j}_{\,\text{cond}},\rmi c \rho_{\text{cond}})\,{,}
\end{equation}
where
\begin{align}
\mathbf{j}_{\,\text{cond}}=\mathbf{j}^{\,0}+\frac{\mathbf{v}}{v^2}(\gamma
-1)(\mathbf{v}\mathbf{j}^{\,0})\,{,}\label{cond-6} \\
c\rho_{\text{cond}}=\frac{\gamma}{c}(\mathbf{v}\mathbf{j}^{\,0})=
\frac{1}{c}(\mathbf{v}\mathbf{j}_{\,\text{cond}})\,{.}
\label{cond-7}
\end{align}
Here we have employed the Lorentz transformation for the
case, when the  velocity $\mathbf{v}$ of the co-moving
frame $K'$ relative to the inertial reference frame $K$ has
arbitrary orientation \cite{Herg}, \cite[Sec.\ 4]{Pauli},
\cite{NYa}, \cite[Chap.\ IV]{Groot}. The spatial coordinate
axes in $K$ and $K'$ are parallel. Taking into account the
explicit form of the conduction current
$\mathbf{j}_{\,\text{cond}}$ \eqref{cond-6}, one can easily
be convinced that in Eq.\ \eqref{cond-7} the product $\gamma
\mathbf{j}^{\,0}$ may be substituted by
$\mathbf{j}_{\,\text{cond}}$.

Application of   the Lorentz transformation specified above
to the 4-vector \eqref{cond-2} yields
\begin{equation}
\label{cond-8}
{j}^{\text{\,conv}}_\mu=\gamma\left (
c\rho^0\frac{\mathbf{v}}{c},\rmi c\rho^{\,0}\right )=
(\rho_{\text{\,conv}}\mathbf{v},\rmi c \rho_{\text{\,conv}})=c\rho^{\,0}u_\mu\,{,}
\end{equation}
where  $\rho_{\,\text{conv}}=\gamma \rho^{\,0}$ is the
density of the convection charge in an arbitrary inertial reference frame,
$\rho_{\,\text{conv}}\mathbf{v}=\mathbf{j}_{\,\text{conv}}$
is the density of the convection current
in this system, and $u_\mu $ is the 4-velocity of the medium
 \eqref{eu}. Multiplying  \eqref{cond-8} by  $u_\mu$, we obtain
\begin{equation}
\label{cond-9}
u_\mu j_\mu^{\,\text{conv}}=-c\rho^{\,0}\,{.}
\end{equation}
On account of condition \eqref{cond-1} the conduction
current is orthogonal to the 4-velocity of the medium
\begin{equation}
\label{cond-10}
u_\mu j_\mu^{\,\text{conv}}=0\,{,}
\end{equation}
therefore the convection current in Eq.\   \eqref{cond-9}
can be substituted by the total current
\begin{equation}
\label{cond-11}
u_\mu j_\mu=-с \rho^{\,0}\,{.}
\end{equation}
Now Eq.\  \eqref{cond-8} for convection current
can be rewritten in the following way
\begin{equation}
\label{cond-12}
 j_\mu^{\,\text{conv}}=-(u_\nu j_\nu)u_\mu\,{.}
\end{equation}
Substituting \eqref{cond-12} in \eqref{cond}, we obtain
\begin{equation}
\label{cond-13}
 j_\mu^{\,\text{cond}}=j_\mu-j_\mu^{\,\text{cond}}=
j_\mu+(u_\nu j_\nu)u_\mu\,{.}
\end{equation}
Obviously the conduction current defined in this way
satisfies condition \eqref{cond-1}.

The derived formulae enable us to deduce the material
relation for current \cite{M-zwei}. In the co-moving frame
the Ohm law reads
\begin{equation}
\label{om}
\mathbf{j}^{\,\prime}_{\,\text{cond}}=\lambda \,\mathbf{E}'{,}
\end{equation}
where $\lambda$ is the conductivity of the medium.

Evidently, the covariant extension of this formula is
given by
\begin{equation}
\label{om-1}
j_\mu^{\,\text{cond}}=\lambda\,E_\mu\,{,}
\end{equation}
where the Lorentz vector $E_\mu$ is defined in
\eqref{Kf-1}.

Using Eqs.\ \eqref{Kf-1} and \eqref{cond-1} one can easily
check that in the rest frame  $K'$ Eq.\  \eqref{om-1}
assumes the form \eqref{om}. The covariant Ohm law \eqref{om-1} can be rewritten
by making use of only the total current \eqref{cond-13}
\begin{equation}
\label{om-2}
j_\mu+(j_\nu u_\nu)u_\mu=\lambda\,E_\mu\,{.}
\end{equation}
Substituting in Eq.\ \eqref{om-1} the explicit expression for
the Lorentz vector $E_\mu$, we obtain
\begin{align}
j_\mu^{\,\text{cond}}&=(\mathbf{j}_{\text{\,cond}},\rmi
c \rho_{\text{cond}})=\lambda\,E_\mu \nonumber \\
&=\lambda \gamma \{
\mathbf{E+[qB]},\rmi (\mathbf{qE})\}\,{,}
\label{om-3}\\
\mathbf{j}_{\text{\,cond}}&=\lambda \gamma (\mathbf{E}+[\mathbf{qB}])\,{,}  \label{om-4} \\
c \rho_{\text{cond}}& =\lambda \gamma (\mathbf{q E})
=(\mathbf{q}\mathbf{j}_{\text{\,cond}})\,{.} \label{om-5}
\end{align}
Equation   \eqref{om-5} was derived previously by
making use of the Lorentz transformation (see Eq.\  \eqref{cond-7}).

In the sequel we shall also use the covariant generalization of
the 3-dimensional Poynting vector
\begin{equation}
\label{P}
\mathbf{S}^{\text{P}}=\frac{c}{4\pi}\,[\mathbf{EH}]\,{.}
\end{equation}
This generalization  was constructed  by Minkowski
\cite[pp.\ 34,35]{M-zwei} with the name  Ruhstrahlvektor
\begin{equation}
\label{R-S}
\Omega_\mu= -\rmi \varepsilon_{\mu \alpha \beta \gamma}E_\alpha H_\beta u_\gamma \,{.}
\end{equation}
One can easily be convinced that in the co-moving reference
frame, where $u_\alpha =(0,0,0,\rmi)$, the vector  \eqref{R-S}
has the components
\begin{equation}
\label{R-S-1}
\Omega'_\mu= ([\mathbf{E'H'}],\,0)\,{.}
\end{equation}
For simplicity we shall call the vector  $\Omega_\mu$ by Minkowski vector.

We shall also use the 4-vector  $\wt_\mu$, which is obtained
from the Minkowski vector $\Omega_\mu$ \eqref{R-S} by the substitution
\begin{equation}
\label{R-S-2}
 \mathbf {E} \to \mathbf
{D},\quad \mathbf {H} \to \mathbf {B}, \quad
\mathbf {D} \to \mathbf {E}, \quad \mathbf {B}
\to \mathbf {H}\,{.}
\end{equation}
On the analogy of \eqref{R-S} the vector  $\wt_\mu$  is defined by the formula
\begin{equation}
\label{R-S-3}
\wt_\mu = -\rmi \varepsilon_{\mu \alpha \beta \gamma}D_\alpha B_\beta u_\gamma \,{.}
\end{equation}
In the co-moving reference frame the 4-vector $\wt_\mu$ has
the components
\begin{equation}
\label{R-S-4}
\wt'_\mu= ([\mathbf{D'B'}],\,0)\,{.}
\end{equation}
Obviously the vectors $\Omega_\mu$ and $\wt_\mu$ satisfy the condition
\begin{equation}
\label{R-S-5}
\Omega_\mu u_\mu=0, \quad \wt _\mu u_\mu=0\,{.}
\end{equation}

Closing this Section, one can say briefly that the constitutive relations take
into account the influence of the media on the
electromagnetic field.\footnote{Media play the role of a
given ``background'' for the dynamics of electromagnetic
field, i.e., it is supposed that the material
characteristics of the media, $\varepsilon$ and $\mu$, do
not depend on  the external electromagnetic field. This
specifies, in particular, the range of applicability of the
macroscopic electrodynamics, namely, external fields should
be much less than the intermolecular fields which determine
the electric and magnetic characteristics of media,
$\varepsilon$ and $\mu$. If such dependence is allowed,
then it implies  that the medium under consideration is
nonlinear.} In order to be complete, the macroscopic
electrodynamics should be able to predict the action of the
external field on the media, because the electromagnetic
field itself is manifested, on the physical level, only
through this action, or more precisely through
ponderomotive forces, which are experienced by charges,
currents, and media.\footnote{It is convenient for us to
single out  macroscopic charges and currents from the material medium.}

\section{Ponderomotive forces and the energy-momentum tensor}
\label{ftei} In macroscopic electrodynamics \cite{LL8}, it
is generally accepted the point of view, formulated first
by Lorentz, according to which the macroscopic equations of
electromagnetic field inside a medium should be obtained as
a result of the averaging, over the physically
infinitesimal volume, of the corresponding microscopic
equations in vacuum with allowance for the point-like
charges and currents forming a medium. In this approach,
all macroscopic forces which are exerted by external
electromagnetic field on macroscopic charges and currents,
are the result of averaging
of the microscopic Lorentz forces acting in vacuum on microscopic charges
$\rho_{\,\text{micro}}$ and $\mathbf{j}_{\,\text{micro}}$:
\begin{equation}
\label{Lor-a}
\overline{\mathbf{f}^{\,\text{L}}_{\,\text{micro}}}=\overline{\rho_{\,\text{micro}}\mathbf{e}}+
\frac{1}{c}\,\overline{\mathbf{[j_{\,\text{micro}}h]}}\,{.}
\end{equation}
A line just above the letter denotes the aforesaid  averaging;
 $\mathbf{e}$ and $\mathbf{h}$ are the  strengthes of
microscopic electric and magnetic fields, respectively.
Microscopic fields include intermolecular fields and
external field. The microscopic charges and currents are
usually divided into bound and free. It is natural to carry
out this separation in the rest frame $K'$:
\begin{align}
\rho'_{\,\text{micro}}&= \rho'_{\,\text{free}}+\rho'_{\,\text{b}}\,{,} \label{micro-1} \\
\mathbf{j}^{\,\prime}_{\,\text{micro}}&=
\mathbf{j}^{\,\prime}_{\,\text{free}}+\mathbf{j}^{\,\prime}_{\,\text{b}}\,{.}
\label{micro-2}
\end{align}
The averaging in Eqs. \eqref{micro-1} and \eqref{micro-2} yields
evidently the following result:
\begin{align}
\overline{\rho'_{\,\text{micro}}}&= \overline{\rho'_{\,\text{free}}}
+\overline{\rho'_{\,\text{b}}}=0+
\rho'_{\,\text{conv}}=\rho^{\,0}\,{,} \label{micro-3} \\
\overline{\mathbf{j}^{\,\prime}_{\,\text{micro}}}&=
\overline{\mathbf{j}^{\,\prime}_{\,\text{free}}}+\overline{\mathbf{j}^{\,\prime}_{\,\text{b}}}=
\overline{\mathbf{j}^{\,\prime}_{\,\text{free}}}+\bm{0}=
\overline{\mathbf{j}^{\,\prime}_{\,\text{cond}}}=\mathbf{j}^{\,0}\,{,} \label{micro-4}
\end{align}
where the quantities $\rho^{\,0}$ and $\mathbf{j}^{\,0}$
were introduced before in defining the macroscopic
convection current $j_\mu^{\,\text{conv}}$ and macroscopic
conduction current  $j_\mu^{\,\text{cond}}$ (see Eqs.\
\eqref{cond}--\eqref{cond-4}).

Let us assume further that the separate multipliers in Eq.\
\eqref{Lor-a} can be averaged independently.\footnote{This
averaging does not pretend to  a completeness and  rigorous
consideration being illustrative only.} Taking into account
definitions $\overline{\mathbf{e}}=\mathbf{E}$,
$\overline{\mathbf{h}}=\mathbf{B}$ and Eqs.\
\eqref{micro-3}, \eqref{micro-4}, we obtain only the
macroscopic Lorentz force applied to macroscopic charges
and currents in the co-moving frame
\begin{equation}
\label{Lor}
(\mathbf{f}^{\,\text{L}}_{\,\text{macro}})'=
\overline{(\mathbf{f}^{\,\text{L}}_{\,\text{micro}})'}=
\rho^{\,0}\mathbf{E}'+\frac{1}{c}\,[\mathbf{j^{\,0}B'}]\,{.}
\end{equation}
The covariant generalization of the macroscopic Lorentz
force \eqref{Lor} is apparently given by
\begin{equation}
\label{Lor-1}
(f^{\,\text{L}}_\alpha)_{\text{macro}}=\frac{1}{c}\, F_{\alpha\beta}j_\beta
=\{\rho\mathbf{E}+\frac{1}{c}\,[\mathbf{jB}], \rmi (\mathbf{qE})\}\,{,}
\end{equation}
where the current  $j_\alpha = \{\mathbf{j},\rmi c\rho \}$
is the sum of the convection current and conduction current (see
Eqs.\  \eqref{cond}, \eqref{cond-5}, \eqref{cond-8}).

At the end of Sec.\ \ref{Min}, the macroscopic Lorentz force
will be derived separately for the convection current and
for the conduction current; in the latter case we will take into account
the covariant generalization of the Ohm law \eqref{om-1}
and Minkowski constitutive relations.

Thus the simple ``naive'' averaging of the microscopic Lorentz
forces does not grasp the mechanical forces exerted by
external electromagnetic field on the medium. More
sophisticated averaging procedures require specification of
the microscopic (molecular or atomic) structure of the
medium and the results of such averaging turn out to be
dependent on the chosen microscopic model (see, for
example, \cite{Groot}).

Taking into account all of this, one has to follow
indirect way.

When the dielectric\footnote{Until the end of this Section,
the medium  is considered to be at rest.} is placed in an
external electrostatic field $\mathbf{E}$, the volume
density of the respective ponderomotive forces
$\mathbf{f}^{(\text{e})}$ is given by the
formula~\cite{Stratton,PPh,Tamm}
\begin{equation}
\label{e1}
\mathbf{f}^{(\text{e})}_{\text{stat}}=\rho\mathbf{E}-\frac{1}{8\pi}E^2 \bm{\nabla}
\varepsilon+\frac{1}{8\pi}\bm{\nabla}\left (
E^2\frac{\partial \varepsilon}{\partial \tau}\,\tau
\right ){,}
\end{equation}
where $\rho$ is the macroscopic charge density  (see the
second equation in \eqref{m-2}), $\tau$ is the medium
density. This formula is derived by calculating the work
accomplished by the electromagnetic field in dielectric by
virtual displacement of the medium \cite{Stratton,PPh,Tamm}.

Fully similar one can derive the ponderomotive force
exerted by the static magnetic field on the  magnetic
\cite{Stratton,PPh,Tamm}
\begin{equation}
\label{e2}
\mathbf{f}^{(\text{m})}_{\text{stat}}=\frac {1}{c}\,[\mathbf{j\,B}]-\frac{1}{8\pi}H^2 \bm{\nabla}
\mu+\frac{1}{8\pi}\bm{\nabla}\left (
H^2\,\frac{\partial \mu}{\partial \tau}\,\tau
\right ){.}
\end{equation}
Here  $\mathbf{j}$ is the macroscopic current density (see
the first equation in \eqref{m-2}).

The last terms in Eqs.\ \eqref{e1} and \eqref{e2} are the
volume densities of electric striction and magnetic
striction forces, that do not contribute to the net force
applied to a macroscopic body. Therefore we will disregard
striction forces in what follows and assume that the volume
density of the ponderomotive force for static field is defined by
\begin{equation}
\label{e3}
\mathbf{f}_{\text{stat}}=\mathbf{f}^{(\text{e})}_{\text{stat}}+
\mathbf{f}^{(\text{m})}_{\text{stat}}=\rho\mathbf{E}+\frac
{1}{c}\,[\mathbf{j\, B}]-\frac{1}{8\pi}E^2
\bm{\nabla} \varepsilon-\frac{1}{8\pi}H^2 \bm{\nabla} \mu
\,{.}
\end{equation}
The first two terms on the right hand side of Eq.\
\eqref{e3} is the macroscopic Lorentz force \eqref{Lor-1},
exerted on charges and currents. The last two terms on the
right hand side of \eqref{e3} define the ponderomotive
forces which are exerted by external electrostatic field
$\mathbf{E}$ and by external magnetic field $\mathbf{H}$ on
the medium. Ultimately we are interested just in these
forces, but in intermediate calculations it is convenient
for us to retain in \eqref{e3} the Lorentz force
\eqref{Lor-1} too.

In the textbooks \cite{Stratton,PPh,Tamm} it is proved that
the ponderomotive force \eqref{e3} caused  by {\it static}
electromagnetic field can be reduced to the Maxwell tensions
\begin{equation}
\label{e4}
\mathbf{f}_{\text{stat}}= \bm{\sigma},\quad
\bm{\sigma}=(\sigma_1,\sigma_2,\sigma_3),\quad {\sigma}_i\equiv\frac{\partial \sigma_{ik}}{\partial x_k} \,{,}
\end{equation}
where $\sigma_{ik}$ is the Maxwell stress tensor of
electromagnetic field in medium
\begin{equation}
\label{e5}
4\pi \sigma_{ik}=E_iD_k+H_iB_k-\frac{\delta_{ik}}{2}\,(\mathbf{ED+HB})
 \,{.}
\end{equation}
In order to verify Eq.\ \eqref{e4}, one has to substitute
Eq.\ \eqref{e5} in \eqref{e4}, take into account the
Maxwell equations \eqref{m-1}--\eqref{m-2} with
$\mathbf{\dot{D}}=0,\; \mathbf{\dot{B}}=0$, and make use of
the constitutive equations \eqref{e0}. As a result Eq.\
\eqref{e4} reduces to \eqref{e3}.

A consistent, physically motivated derivation of the
formula for ponderomotive forces in macroscopic
electrodynamics is restricted to the static external
fields. To extend the obtained expression to the case of
time dependent fields and cast it into explicitly covariant
form  one has to resort to some assumptions and use the
analogy with electrodynamics in
vacuum.\footnote{Difficulties emerged in deriving  the
ponderomotive forces  exerted on  a medium and caused by
time-dependent external electromagnetic field are
substantiated. The point is the static electric and
magnetic fields can be, in principle, treated proceeding
from the conception of action at a distance (instantaneous
transmission of interaction). In contrast to this, the
time-dependent fields result in a finite velocity of
propagation of electromagnetic interaction. Here the idea
of a short-range interaction is adopted, which is based on
the notion of a field playing the role of a physical
carrier of interaction.}

Following this way, we assume  preliminary that the density
of ponderomotive force, $\mathbf{f}$, equals $\bm{\sigma}$
also in the case of time-dependent external electromagnetic
field.  We substitute again \eqref{e5} into \eqref{e4} and
take advantage of the complete Maxwell equations
\eqref{m-1}--\eqref{m-2} and the constitutive relations
\eqref{e0}. This yields
\begin{equation}
\label{e6}
\mathbf{f}=\bm{\sigma}
=\rho\mathbf{E}+\frac
{1}{c}\,[\mathbf{j\, B}]-\frac{1}{8\pi}E^2 \bm{\nabla}
\varepsilon-\frac{1}{8\pi}H^2 \bm{\nabla}
\mu +\frac{\partial}{\partial t}\left (
\frac{1}{4\pi c}\,[\mathbf{DB}]
\right){.}
\end{equation}
The obtained formula contains an additional, in comparison
with \eqref{e3}, term
\begin{equation}
\label{e7}
\frac{\partial}{\partial t}\left (
\frac{1}{4\pi c}\,[\mathbf{DB}]
\right)
\end{equation}
caused by the time-dependence of electromagnetic field. In
the case of vacuum the right-hand side of \eqref{e6} should
obviously contain only the Lorentz force \eqref{Lor}.

It is instructive to retrace in what way the additional
term \eqref{e7} vanishes in vacuum. In this case formula  \eqref{e7}
acquires the form
\begin{equation}
\label{e8}
\frac{\partial}{\partial t}\left (
\frac{1}{4\pi c}\,[\mathbf{EH}]
\right){.}
\end{equation}
In order to elucidate this question we take advantage of
the following statement which is proved in electrodynamics
in vacuum (see the Appendix \ref{A} and also
\cite[\S~30]{Pauli}, \cite[\S~33]{LL2}, \cite{Jackson}):
the Lorentz force \eqref{Lor}, exerted on charges and
currents in vacuum, can be represented in the following
form:
\begin{equation}
\label{e9}
\mathbf{f}^{L}_{\text{vac}}=\bm{\sigma}_{\text{vac}}-\mathbf{\dot{g}}_{\text{vac}}\,{.}
\end{equation}
Here  $\bm{\sigma}_{\text{vac}}$ is  $\bm{\sigma}$
calculated in vacuum, i.e., setting $\mathbf{D=E}$ and
$\mathbf{B=H}$ (see Eqs.\ \eqref{e4}, \eqref{e5}), and
$\mathbf{g}_{\text{vac}}$ is the momentum density of
electromagnetic field in vacuum
\begin{equation}
\label{e10}
\mathbf{g}_{\text{vac}}=\frac{1}{4\pi c}[\mathbf{EH}]\,{.}
\end{equation}
When calculating the right-hand side of Eq.\ \eqref{e9},  the
complete Maxwell equations in vacuum
\eqref{m-1}--\eqref{m-2} should be taken into account (see
the Appendix \ref{A}).

In view of all this, the volume density of ponderomotive
force, Eq.\  \eqref{e6}, exerted by the time-dependent
external electromagnetic field on a medium, is generalized in the
following way:
\begin{equation}
\label{e11}
\mathbf{f}= \bm{\sigma}-\mathbf{\dot{g}}\,{,}
\end{equation}
where  $\mathbf{g}$ is the  linear momentum density of
electromagnetic field in medium.

Unfortunately there is no generally accepted expression for
the electromagnetic momentum density in a medium
$\mathbf{g}$. In this field, fruitless  argument  is
already going on more a century. In most cases, the
attention is payed to the definitions of $\mathbf{g}$
proposed by Minkowski and by Abraham in their time
\cite[Sec.~35]{Pauli}. It is these versions that will be
considered in the present paper.

\section{Ponderomotive forces in the Minkowski approach}
\label{Min} Minkowski \cite{M-zwei} defined electromagnetic
momentum density in a medium by the formula
\begin{equation}
\label{e13}
\mathbf{g}^{\text{M}}= \frac{1}{4\pi c}\,[\mathbf{DB}]\,{.}
\end{equation}
This choice leads, in particular, to cancellation of the
additional term \eqref{e7} in Eq.\ \eqref{e11}. As a
result, Eq.\ \eqref{e11} defining the ponderomotive force
density in the case of time-dependent electromagnetic
field, coincides, after expanding $\bm{\sigma}$ and
$\mathbf {\dot{g}}$, with \eqref{e3} which defines this
force for static external field
\begin{align}
\left .\mathbf{f}^{\text{M}}\right |_{\mathbf{v}=0}&=\bm{\sigma} -
 \mathbf{\dot g}^{\text{M}}{} \label{e-14a}\\ &=
\rho \mathbf{E}+\frac{1}{c}\,[\mathbf{jB}]-
\frac{1}{8 \pi} \mathbf{E}^2 \bm{\nabla}\varepsilon -
\frac{1}{8 \pi} \mathbf{H}^2 \bm{\nabla}\mu= \mathbf{f}_{\text{stat}}\,{.}
\label{e-14}
\end{align}
It will be recalled that a medium at rest is considered and
in deriving Eq.\ \eqref{e-14} constitutive relations
\eqref{e0} have been used.  Hence Eq.\ \eqref{e-14} is
defined on $\g$.

In order to recast Eq.\ \eqref{e-14a} into  covariant form
one has obviously to extend the Maxwell 3-dimensional
stress tensor \eqref{e5} to the 4-dimensional
stress-energy-momentum tensor. Minkowski \cite{M-zwei}
defined the energy-momentum tensor of electromagnetic
field in a medium in the following form:
\begin{equation}
\label{tm-cov} T^{\text{M}}_{\alpha
\beta}=\frac{1}{4\pi}\left ( F_{\alpha
\gamma}H_{\gamma \beta}-\frac{\delta_{\alpha
\beta}}{4} F_{\gamma \nu }H_{\nu \gamma} \right
){.}
\end{equation}
By analogy with  electrodynamics in vacuum (see Eq.\
\eqref{a5}), formula  \eqref{e-14a}, in a covariant form, is written as
\begin{equation}
\label{ftm}
f^{\text{M}}_\mu =\frac{\partial T^{\text{M}}_{\mu \nu}}{\partial x_\nu}\,{.}
\end{equation}
The individual  components of the tensor $T^{\text{M}}_{\alpha
\beta}$, defined in \eqref{tm-cov}, are given by the formulae
\begin{equation}
\label{tm-1} \tm=\left (
\begin{matrix}
\sigma^{\text{M}}_{ij}&-\rmi c\,
\mathbf{g}^{\text{M}}\cr -\frac{\displaystyle
\rmi}{\displaystyle  c}
\,\mathbf{S}^{\text{M}}&w^{\text{M}}\cr
\end{matrix}
\right ){,}
\end{equation}
where
\begin{gather}
\sigma_{ij}^{\text{M}}=\frac{1}{4 \pi}\left \{
E_iD_j+H_iB_j-\frac{\delta_{ij}}{2}\,(\mathbf{ED+HB})
\right \}{,}\label{tm-2} \\
\frac{1}{c}\,
 \mathbf {S}^{\text{M}}=\frac{1}{c}\,\mathbf{S}^{\text{P}}
=\frac{1}{4\pi}\,[\mathbf{EH}]\,{,}\quad c\mathbf {g}^{\text{M}}=
\frac{1}{4\pi}\,[\mathbf{DB}]\,{,}
\label{tm-3} \\
w^{\text{M}}=\frac{1}{8
\pi}\,(\mathbf{ED+HB})\,{.}\label{tm-4}
\end{gather}
The Minkowski stress tensor \eqref{tm-2} coincides with the
Maxwell expression \eqref{tm-2}, the density of the energy
current in the Minkowski energy-momentum tensor
\eqref{tm-3} is given by the Poynting vector
$\mathbf{S}^{\text{P}}$ (see Eq.\ \eqref{P}), and the
momentum density $\mathbf {g}^{\text{M}}$ in \eqref{tm-3}
reproduces Eq.\ \eqref{e13}.

By definition,  Eq.\ \eqref{tm-cov}   determines the
Minkowski energy-momentum tensor both for  media at rest
and for moving media. The velocity of a medium,
$\mathbf{v}$, does not enter into Eq.\ \eqref{tm-cov}. The
Minkowski tensor  \eqref{tm-cov} is not symmetric
$T^{\text{M}}_{\alpha \beta}\neq T^{\text{M}}_{\beta
\alpha}$, its trace vanishes  $T^{\text{M}}_{\alpha
\alpha}=0$.

Now we turn to  calculating  the explicit covariant formula
for ponderomotive force by making use of the covariant formula
\eqref{ftm}.  For simplicity sake we introduce the notation
\begin{equation}
\label{partial}
\frac{\partial \varphi}{\partial x_\nu}\equiv\varphi_{,\nu}\,{.}
\end{equation}
On substituting
\eqref{tm-cov} into \eqref{ftm}, we get
\begin{align}
4 \pi T^{\text{M}}_{\alpha \beta,\beta}&=
 F_{\alpha \gamma,\beta}H_{\gamma
\beta}+F_{\alpha\gamma}H_{\gamma\beta,\beta}-\frac{\delta_{\alpha \beta}}{4}\left (
F_{\gamma \nu}H_{\nu \gamma}\right )_{,\beta} \nonumber
\\&=F_{\alpha\gamma,\beta}H_{\gamma \beta}+\frac{4 \pi}{c}\,
F_{\alpha\gamma}j_{\gamma}-\frac{1}{4}\, (F_{\gamma
\nu}H_{\nu \gamma})_{,\alpha}\,{.} \label{B-2}
\end{align}
Here we have taken advantage of nonhomogeneous Maxwell
equations \eqref{m-6}. Further we transform the first term
on the right-hand side in the last equality in \eqref{B-2}
\begin{align}
F_{\alpha\gamma ,\beta}H_{\gamma
\beta}=\frac{1}{2}\,(F_{\alpha\gamma
,\beta}H_{\gamma\beta}+F_{\alpha\beta
,\gamma}H_{\beta\gamma}){}\nonumber \\=-\frac{1}{2}\,(F_{\gamma\alpha,\beta}+F_{\alpha\beta,\gamma})H_{\gamma\beta}=
&\frac{1}{2}\,F_{\beta\gamma,\alpha}H_{\gamma\beta}\,{.}
\label{B-3}
\end{align}
When deriving the last equality in \eqref{B-3} we have used
the homogeneous Maxwell equations \eqref{m-5}. The
substitution of \eqref{m-5} into \eqref{B-2} yields
\begin{align}
4 \pi T^{\text{M}}_{\alpha \beta,\beta}&=\frac{4
\pi}{c}\,F_{\alpha \gamma}j_{\gamma
}+\frac{1}{2}(F_{\beta\gamma}H_{\gamma\beta})_{,\alpha}-\frac{1}{2}\,
F_{\beta \gamma }H_{\gamma \beta,\alpha}-\frac{1}{4}\,
(F_{\gamma\nu}H_{\nu\gamma})_{,\alpha}\nonumber\\
&=\frac{4 \pi}{c}\,F_{\alpha
\gamma}j_{\gamma}+\frac{1}{4}\,
(F_{\beta\gamma}H_{\gamma\beta})_{,\alpha}-\frac{1}{2}\,F_{\beta\gamma}H_{\gamma
\beta,\alpha}
\nonumber
 \\ &=\frac{4 \pi}{c}\,F_{\alpha
\gamma}j_{\gamma}+\frac{1}{4}\,(F_{\beta\gamma,\alpha}H_{\gamma\beta}-
F_{\beta\gamma}H_{\gamma\beta,\alpha})\,{.} \label{B-4}
\end{align}
In another way formula \eqref{B-4} was derived in Ref.\
\cite[p.\ 43]{M-zwei}.

Now we transform the expression in round brackets on the
right-hand side of Eq.\ \eqref{B-4} by making use of the
Minkowski constitutive relations \eqref{Min-1},
\eqref{Min-2}, i.e., on $\g$. The definitions \eqref{Kf-1}
and \eqref{Kf-2} can be inverted, expressing  the tensors
$F_{\mu \nu}$, and $H_{\mu \nu}$ in terms of $E_\mu,
B_\mu,D_\mu,H_\mu$ and $u_\mu$:
\begin{align}
F_{\mu\nu}&=u_\mu E_\nu-u_\nu E_\mu+\rmi
\varepsilon_{\mu\nu\gamma\rho}u_\gamma B_\rho\,{,} \label{Kf-1r} \\
H_{\mu\nu}&=u_\mu D_\nu-u_\nu D_\mu+
\rmi \varepsilon_{\mu\nu\gamma\rho}u_\gamma H_\rho \,{.} \label{Kf-2r}
\end{align}
These relations can be easily checked in the co-moving
reference frame $K'$, where $u'_\gamma =(0,0,0,\rmi)$, and
by virtue of their covariance they are valid in an
arbitrary inertial reference frame $K$.

In \eqref{Kf-1r} and \eqref{Kf-2r} we take into account
the constitutive relations \eqref{Min-1} and \eqref{Min-2}
\begin{align}
F_{\mu\nu}&=u_\mu E_\nu-u_\nu E_\mu+\rmi \mu
\varepsilon_{\mu\nu\gamma\rho}u_\gamma H_\rho \nonumber \\
&= {\cal E}_{\mu \nu}+\rmi \mu {\cal H}_{\mu \nu}\,{,} \label{Kf-1rr} \\
H_{\mu\nu}&=\varepsilon (u_\mu E_\nu-u_\nu E_\mu)+
\rmi \varepsilon_{\mu\nu\gamma\rho}u_\gamma H_\rho \nonumber \\
&= \varepsilon {\cal E}_{\mu \nu}+\rmi {\cal H}_{\mu \nu}\,{.} \label{Kf-2rr}
\end{align}
Here the notations are introduced
\begin{gather}
 {\cal E}_{\mu \nu}=- {\cal E}_{ \nu \mu}=u_\mu E_\nu-u_\nu E_\mu,\quad
{\cal E}_{\mu \nu}{\cal E}_{\nu \mu }=2E^2_\nu\,{,} \nonumber \\
 {\cal H}_{\mu \nu}= - {\cal H}_{\nu \mu} =\varepsilon_{\mu \nu \gamma \rho}u_\gamma H_\rho,
\quad {\cal H}_{\mu \nu}{\cal H}_{\nu \mu }=2H^2_\rho,\quad  {\cal E}_{\mu \nu} {\cal H}_{\nu\mu }=0\,{.}
\label{EH}
\end{gather}
On substituting  \eqref{Kf-1rr} and \eqref{Kf-2rr} into
expression in round brackets in Eq.\  \eqref{B-4} we get
\begin{multline}
F_{\beta\gamma,\alpha}H_{\gamma\beta}-F_{\beta\gamma}H_{\gamma\beta,\alpha}={}
 \\
{}= ({\cal E}_{\beta\gamma}+\rmi \mu {\cal H}_{\beta\gamma})_{,\alpha}
(\varepsilon{\cal E}
_{\gamma \beta}+\rmi {\cal H}_{\gamma\beta})-
({\cal E}_{\beta\gamma}+\rmi \mu {\cal H}_{\beta\gamma})(\varepsilon{\cal E}
_{\gamma \beta}+\rmi {\cal H}_{\gamma\beta})_{,\alpha } ={}\\
{}=-\varepsilon_{,\alpha}
{\cal E}_{\gamma\beta}{\cal E}_{\beta\gamma}-
\mu_{,\alpha}{\cal H}_{\gamma\beta}{\cal H}_{\beta\gamma}+N_\alpha=
-2\varepsilon_{,\alpha}E^2_\nu-2\mu_{,\alpha}H^2_\rho+N_\alpha \,{,}
\label{B-5}
\end{multline}
where the vector  $N_\alpha$ is
\begin{equation}
N_\alpha=\rmi(\varepsilon\mu-1)({\cal E}_{\beta\gamma} {\cal H}_{\gamma\beta,\alpha}
-{\cal E}_{\beta\gamma,\alpha} {\cal H}_{\gamma\beta})\,{.}
\label{N}
\end{equation}
Obviously  $N_\alpha=0$ if the velocity $u_\beta$ is the
same at any points of a medium, i.e., $u_{\beta,\alpha}=0$.
Indeed, in this case the velocity $u_\beta$ in \eqref{N}
can be taken out of the differentiation sign and each term
will contain the multiplier $\varepsilon_{\mu \nu \rho \sigma}u_\nu
u_\rho \ldots$ which vanishes.

Now let the medium velocity be varying in space and in
time. In this case a nonvanishing contribution  to
\eqref{N} arises only when  the differentiation operator
$\partial_\alpha$ acts on the medium velocity but not on
the vectors $E_\nu $ and  ${H_\mu}$ (see definitions of
${\cal E}_{\mu \nu}$ and  ${\cal H}_{\mu \nu}$  in
\eqref{EH}). Taking into account all this,  one finds easily
\begin{equation}
{\cal E_{\beta\gamma}} {\cal
H_{\gamma\beta,\alpha}}=-2\,\rmi\, \Omega_\nu
u_{\nu,\alpha}, \quad {\cal E_{\beta\gamma,\alpha}} {\cal
H_{\gamma\beta}}=-2\,\rmi\, \Omega_\nu
u_{\nu,\alpha}\,{,}\label{N-1}
\end{equation}
where $\Omega_\nu$ is the Minkowski vector \eqref{R-S}.
Substitution of \eqref{N-1} into \eqref{N} gives
\begin{equation}
N_\alpha=4(\varepsilon\mu-1)\Omega_\nu u_{\nu,\alpha}\,{.}
\label{N-2}
\end{equation}
Further simplification of Eq. \eqref{N-2} can be achieved
by expressing its right-hand side in terms of spatial
components of the vectors $\Omega_\nu$ and  $u_\nu$ by
making use of Eqs.\ \eqref{eu} and  \eqref{R-S-5}. To this
end we introduce the 3-dimensional vector $\W$ through the
equality~\cite{NN-A}
\begin{equation}
(\varepsilon\mu-1)\,\Omega_\alpha =\left (
\frac{\W}{\gamma},\frac{\rmi}{\gamma}(\mathbf{q}\W)
\right ){.}
\label{N-3}
\end{equation}
In this definition the  factor $\gamma^{-1}$ is introduced
which is canceled in subsequent equations where
$(\varepsilon \mu-1)\Omega_\alpha$ is multiplied by
$u_\beta$. Now Eq.\  \eqref{N-2} assumes the form
\begin{equation}
N_\alpha=4 \left (
\W\mathbf{q}_{,\alpha}
\right ){.}
\label{N-4}
\end{equation}
On the physical configuration space $\g$, the vector ${\W}$
is given by (see the Appendix \ref{B})
\begin{equation}
\W=[\mathbf{DB}]-[\mathbf{EH}]+\gamma^2 \mathbf{q(q,[DB]-[EH])}\,{.}
\label{N-5}
\end{equation}
It is worth noting that the medium characteristics
$\varepsilon$ and  $ \mu$ do not enter into this formula
explicitly. Substituting \eqref{B-5} and  \eqref{N-4} in
\eqref{B-4}, we obtain on $\g$ the final compact formula for
the ponderomotive forces in the Minkowski approach
\begin{equation}
f_\alpha^{\text{M}}\equiv T^{\text{M}}_{\alpha
\beta,\beta}=\frac{1}{c}\,F_{\alpha\gamma}j_\gamma- \frac{1}{8\pi}\left (
\varepsilon_{,\alpha}E^2_\nu +\mu _{,\alpha}H_\nu^2
 -2(\W\mathbf{q}_{,\alpha})\right )\,{,}
\label{N-6}
\end{equation}
where the Lorentz vectors  $E_\nu$ and  $H_\nu $ are
defined  in \eqref{Kf-1} and \eqref{Kf-2}.

Let us write down separately the spatial and  temporal
components of the 4-force $f_\alpha^{\text{M}}$ defined in
\eqref{N-6},
\begin{align}
\mathbf{f}^{\text{M}}&=\rho \mathbf{E}+\frac{1}{c}\, [\mathbf{jB}]-
\frac{1}{8\pi}\left(
E_\nu^2 \bm{\nabla} \varepsilon+H^2_\nu \bm{\nabla} \mu-2(\frak{W}_i\bm{\nabla} q_i)
\right){,}
\label{N-7} \\
\frac{c}{\rmi} \,f_4^{\text{M}} &=(\mathbf{jE})+\frac{1}{8\pi}\,\left (
\dot {\varepsilon}E^2_\nu +\dot{\mu}H^2_\nu-2 (\W\mathbf{\dot q})
\right ){.} \label{N-8}
\end{align}
Differentiation with respect to time in \eqref{N-8} is
denoted, as before, by a dot. If the medium is at rest
$(\mathbf{v}=0)$, then the force  $\mathbf{f}^{\text{M}}$
is converted  into Eq.\ \eqref{e-14}, which in turn
coincides with Eq.\ \eqref{e3} for the ponderomotive force
exerted by a {\it static} external electromagnetic field.
Let us note  the peculiarities of the obtained expression
\eqref{N-7} for the ponderomotive forces exerted on the
medium in the Minkowski approach.
\begin{itemize}
\item This formula does not contain the derivatives of the
field vectors $\mathbf{E,\,H,\,D}$, and $\mathbf{B}$.
\item If the medium is homogeneous ($\varepsilon_{,k}=0,\;
\mu_{,k}=0$) and its velocity is the same at all medium
points ($\mathbf{q}_{,k}=0$), then the ponderomotive force
vanishes.
\end{itemize}

It should be emphasized that we succeeded in revealing  these
features of the force $\mathbf{f}^{\,\text{M}}$ just
because the calculation was carried out on the physical
configuration space $\g$.

To make presentation complete, we write down, in an explicit
form, the Lorentz force (the first term on the right-hand
side of Eq.\ \eqref{N-6}) for special cases when there is
only convection current (the respective Lorentz force is
$f_\alpha^{\,\text{conv}}$) or only the conduction current
(the Lorentz force is $f_\alpha^{\,\text{cond}}$).

In the first case, with allowance for Eq.\ \eqref{cond-8}, we obtain
\begin{equation}
\label{N-9}
f_\alpha^{\,\text{conv}}=\frac{1}{c}\, F_{\alpha\gamma}j_\gamma^{\,\text{conv}}
=\rho^0F_{\alpha\gamma}u_\gamma=\rho^0 E_\alpha\,{,}
\end{equation}
where $\rho^0 $ is the density of convection charges in the
co-moving reference frame \eqref{cond-2}, \eqref{cond-3}.

In deriving  the Lorentz force  experienced by the
conduction current we take first into account the
covariant Ohm law \eqref{om-1}
\begin{equation}
\label{N-10}
f_\alpha^{\,\text{cond}}=\frac{1}{c}\, F_{\alpha\beta}j_\beta^{\,\text{cond}}
=\frac{\lambda}{c}\,F_{\alpha\beta}E_\beta\,{.}
\end{equation}
And now we take advantage of the representation
\eqref{Kf-1rr} for the tensor $F_{\alpha \beta}$, which
holds on~$\g$,
\begin{align}
f_\alpha^{\,\text{cond}}&=\frac{\lambda}{c}\,(u_\alpha
E_\beta-u_\beta E_\alpha)E_\beta+ \rmi
\mu\frac{\lambda}{c}\,\varepsilon_{\alpha\beta\gamma\delta}u_\gamma
H_\delta E_\beta \nonumber \\
&=\frac{\lambda}{c}\,u_\alpha E^2_\gamma+\mu\frac{\lambda}{c}\,\Omega_\alpha \,{,}
\label{N-11}
\end{align}
where $\Omega_\alpha$ is  the Minkowski vector \eqref{R-S}.
In the rest frame we have
\begin{equation}
\label{N-12}
(\mathbf{f}^{\text{\,cond}})'=
\mu \frac{\lambda}{c}\,[\mathbf{E'H'}], \quad (f_4^{\,\text{cond}})'=\rmi \frac{\lambda}{c}\,\mathbf{E'}^2
\,{.}\end{equation}
\section{Different representations of the Abraham energy-momentum tensor}
\label{Ab-dif} The calculation of the  ponderomotive forces
in the framework of the Abraham approach is simplified
essentially if a special representation of the Abraham
energy-momentum  tensor is employed. Let us consider briefly
different representations  for this tensor~\cite{NN-A}.

Abraham was guided by the symmetry requirement imposed on
the energy-momentum tensor in a medium. He has constructed
his tensor  taking into account the Minkowski constitutive
relations, i.e., on $\g$~\cite{AP1909,AP1910,Abraham1920}.

Considerably later the explicitly covariant representation
for the symmetric energy-momentum tensor was found which
holds on the whole configuration space $\Gamma$
\cite{Schmutzer}, \cite{Groot}, \cite{Obukhov,MR}, \cite{NN-A}. The
equivalence on $\g$ of the covariant representation  and
the original 3-dimensional Abraham formulae was proved in
Ref.\ \cite{NN-A}.

Here we derive the original  Abraham formulae, which will
be used further, proceeding from the  covariant representation for
the energy-momentum tensor. The Abraham derivation of the
energy-momentum tensor is traced in detail in Ref.\
\cite{NN-A}.

The explicitly covariant  symmetric energy-momentum tensor
of electromagnetic field in a medium $\tsym$ is defined by
the formulae~\cite{NN-A}:
\begin{equation}
\label{sym}
\tsym=\frac{1}{2}\,(\tm +T^{\text{\,M}}_{\beta \alpha})+A_{\alpha \beta}{,}
\end{equation}
where
\begin{gather}
A_{\alpha \beta}=A_{\beta \alpha}=\frac{1}{8\pi}\,
\{u_\alpha(\wt_\beta-\Omega_\beta)+u_\beta(\wt_\alpha-\Omega_\alpha)\} \label{sym-1}\\
=\frac{1}{8\pi}\,\{
u_\alpha(F_{\beta \nu}H_{\nu \lambda}u_\lambda-H_{\beta\nu}F_{\nu\lambda}u_\lambda)+
u_\beta(F_{\alpha \nu}H_{\nu \lambda}u_\lambda-H_{\alpha\nu}F_{\nu\lambda}u_\lambda)
\}\,{.}\label{sym-2}
\end{gather}
Here $\Omega_\alpha$ and  $\widetilde \Omega_\beta$ are the
Minkowski vectors  \eqref{R-S} and \eqref{R-S-3},
respectively. By making use of the values of these vectors
in the co-moving reference frame $K'$ (Eqs.\ \eqref{R-S-1} and
\eqref{R-S-4}) one can easily  show that the tensor~\eqref{sym}
\begin{equation}
\label{sym-3} T_{\alpha \beta}^{\,\text{sym}}=\left (
\begin{matrix}
\sigma_{ij}^{\,\text{sym}}&-\rmi c\,
\mathbf{g}^{\,\text{sym}}\cr -\frac{\displaystyle
\rmi}{\displaystyle  c}
\,\mathbf{S}^{\,\text{sym}}&w^{\text{sym}} \cr
\end{matrix}
\right )
\end{equation}
in $K'$ has the components
\begin{gather}
\sigma^{\prime}_{ij}=\frac{1}{2}\left (
\sigma'^{\,\text{M}}_{ij}+ \sigma'^{\,\text{M}}_{ji}
\right ){,}\label{sym-4} \\
\frac{1}{c}\,
 \mathbf
{S}^{\prime\,\text{sym}}=c\mathbf {g'}^{\,\text{sym}}= \frac{1}{c}\,\mathbf{S}^{\prime\,\text{P}}
=\frac{1}{4\pi}\, [\mathbf{E'H'}]\,{,}
\label{sym-5} \\
w'^{\,\text{sym}}=w'^{\,\text{M}}=\frac{1}{8
\pi}\,(\mathbf{E'D'+H'B'})\,{.}\label{sym-6}
\end{gather}
It will be recalled that  we mark by prime the quantities in
the co-moving reference frame $K'$. Conditions
\eqref{sym-4}--\eqref{sym-6} are satisfied without invoking
constitutive relations \eqref{e0}. It is this point that
distinguish the tensor \eqref{sym} from the energy-momentum
tensor constructed by Abraham (see further and Ref.\
\cite{NN-A}). Conditions \eqref{sym-4}--\eqref{sym-6}
specify the tensor  $\tsym$ uniquely.

Equations  \eqref{sym}--\eqref{sym-2} were derived in our
paper \cite{NN-A} by generalization of the Abraham
reasoning. Now, going in the opposite direction, we show
how to obtain the Abraham formulae proceeding from the
covariant representation \eqref{sym}--\eqref{sym-2}.  In
this consideration Eq.\  \eqref{sym-1} plays a key role
(this formula was derived in Ref.\  \cite{NN-A}).

The tensor $\tsym$ \eqref{sym} is defined on the whole
configuration space $\Gamma $ while Abraham has derived his
formulae with allowance for the constitutive relations,
i.e., on $\g$. Thus we have to take into account the
Minkowski material relations \eqref{Min-1}, \eqref{Min-2}
in Eqs.\ \eqref{sym}--\eqref{sym-2}. It is worth noting
that Abraham used the Minkowski constitutive relations
specifically. Namely, they were not employed to express one
pair of field vectors $\mathbf{E,\,H,\,D}$, and $\mathbf{B}$
in terms of the rest pair, but a special formulae were used
which contain all the vectors $\mathbf{E,\,H,\,D,\,B,\,v}$,
and the material characteristics of  the medium
$\varepsilon$ and $\mu$, these formulae being valid only
under fulfilment of the  Minkowski constitutive relations,
i.e., on $\g$. An important example of such formulae is
the  equality \eqref{w-4}, proved in the Appendix \ref{B}, and
the following  relations which is also valid on $\g$:
\begin{equation}
\label{o-o} \wt_\alpha-\Omega_\alpha=(\varepsilon \mu -1)\,
\Omega_\alpha\,{.}
\end{equation}
Equation \eqref{o-o} can be easily derived if we substitute
in the definition of $\wt_\alpha$ (Eq.\  \eqref{R-S-3}) the
constitutive relations in the form \eqref{Min-1},
\eqref{Min-2} and take into account the definition of
$\Omega_\alpha$ in Eq.\  \eqref{R-S}. As a result the
tensor \eqref{sym}, \eqref{sym-1} reduces on $\g$
to the Abraham energy-momentum tensor~\cite{NN-A}:
\begin{equation}
\label{tA1}
T_{\alpha \beta}^{\,\text{A1}}=\frac{1}{2}\, (\tm +T^{\,\text{M}}_{\alpha \beta})+
\frac{\varepsilon \mu-1}{8\pi}(u_\alpha \Omega_\beta-u_\beta \Omega_\alpha)\,{.}
\end{equation}

The Abraham tensor $T^{\text{A1}}_{\alpha \beta}$ in the
co-moving reference frame $K'$ satisfies conditions
\eqref{sym-4}--\eqref{sym-6} only if the constitutive
relations \eqref{e0} are met, i.e., on $\g$. Therefore
these conditions do not fix this tensor uniquely and it can
be represented on $\g$ in different forms. To show this we
take advantage of the equality \eqref{w-4}, proved in
Appendix \ref{B}. Let us write down the tensor $T_{\alpha
\beta}^{\,\text{A1}}$ and the equality \eqref{w-4},
multiplied by $1/2$, on the neighbouring lines:
\begin{align}
4\pi T^{\,\text{A}1}_{\alpha \beta} &=\frac12\,F_{\alpha
\gamma}H_{\gamma \beta}+\frac12\,F_{\beta \gamma}H_{\gamma
\alpha} -\frac14\,\delta_{\alpha\beta}F_{\gamma
\delta}H_{\delta\gamma}+\frac{\varepsilon\mu-1}{2}\,
(u_\alpha\Omega_\beta+u_\beta\Omega_\alpha)\,{,}\label{tA1a}
\\
0&=\frac12\,F_{\alpha \gamma}H_{\gamma
\beta}-\frac12\,F_{\beta \gamma}H_{\gamma \alpha}
\hbox to
32mm{\hfil$-$\hfil}  \frac{\varepsilon\mu-1}{2}\,
(u_\alpha\Omega_\beta-u_\beta\Omega_\alpha)\,{.}\label{w-4a}
\end{align}
Adding and subtracting the left-hand sides and the
right-hand sides of these equations we obtain on~$\g$\,:
\begin{equation}
\label{ttt}
T^{\,\text{A}1}_{\alpha\beta}=T^{\text{A}2}_{\alpha\beta}=T^{\text{A}3}_{\alpha\beta}\,{,}
\end{equation}
where
\begin{align}
T^{\,\text{A}2}_{\alpha\beta}=\tm +\frac{\varepsilon\mu-1}{4\pi}\,u_\beta\Omega_\alpha\,{,}
\label{tA2}\\
T^{\,\text{A}3}_{\alpha\beta}=T^{\text{M}}_{\beta\alpha}+\frac{\varepsilon\mu-1}
{4\pi}\,u_\alpha\Omega_\beta\,{.}
\label{tA3}
\end{align}
Following in this way,  one can also obtain other
representations  for the Abraham energy-momentum tensor on
$\g$~\cite{NN-A}. The introduction of the vector $\W$ (Eq.\
\eqref{N-5}) through the substitution \eqref{N-3} removes
the explicit  dependence on $\varepsilon \mu$ in tensors
$T_{\alpha \beta}^{\,\text{A}i},\; i=1,2,3$. By making use
of the vector $\W$, Abraham derived explicit formulae, in
terms of the 3-dimensional vectors, for all the components
of the tensors \eqref{tA1} and \eqref{tA2} (see
respectively Refs.\ \cite{AP1910} and \cite{Abraham1920} as
well as our paper \cite{NN-A}). However we shall not use
these 3-dimensional vector formulae further. In calculation
of the ponderomotive forces in the Abraham approach we will
employ  Eq.\ \eqref{tA2}.

It is worthy to note that Eqs.\ \eqref{tA1}, \eqref{tA2},
and \eqref{tA3} for the Abraham tensor were constructed by
Grammel \cite{Grammel}. Possibility of representing the
Abraham tensor in different forms, not all of them being
explicitly symmetric (see Eqs.\ \eqref{tA2} and
\eqref{tA3}), is explained by a restricted domain of their
validity, namely, they hold only on~$\g$.

\section{Ponderomotive forces in the Abraham approach}
\label{Ab-f} At the outset we envisage  the medium at rest
and take advantage of Eq.\ \eqref{e11}. Following the
requirement of the symmetry of the energy-momentum tensor,
Abraham defined (in fact postulated) the momentum density
of electromagnetic field in a medium in the form
\begin{equation}
\label{ga}
\mathbf{g}^{\text{A}}=\frac{1}{4\pi c}\,[\mathbf{EH}]
\end{equation}
(see Eq.\ \eqref{sym-5}). As a result, the additional term
\eqref{e7} in Eq.\ \eqref{e11} is not canceled  and we get
\begin{align}
\mathbf{f}^{\text{A}}|_{\mathbf{v}=0}&=\bm{\sigma} -\mathbf{\dot g}^{\text{A}} \nonumber \\
&=\mathbf{f}^{\text{M}}|_{\mathbf{v}=0}+\frac{1}{4\pi c}\,\frac{\partial}{\partial t}
(\mathbf{[DB]-[EH]})\,{.} \label{ga-1}
\end{align}
The last term in this formula is the Abraham
force \cite[\S~75]{LL8}. We introduce a special notation
for this term
\begin{equation}
\label{ga-2} \Delta
\mathbf{f}^{\text{A}}|_{\mathbf{v}=0}=\frac{1}{4\pi c}
\,\frac{\partial }{\partial t}\,(\mathbf{[DB]-[EH]})\,{,}
\end{equation}
 in oder to distinguish this force from the total
ponderomotive force $\mathbf{f}^{\text{A}}$ in the Abraham
approach.

Now we turn to the derivation of the covariant formula for
the ponderomotive 4-force in the Abraham approach in the
case of moving medium. Abraham, just as Minkowski, lets the
ponderomotive force to be equal  to the divergence  of the
energy-momentum tensor of electromagnetic field in a
medium. For our consideration  it is crucial that the
Abraham energy-momentum tensor  can be represented on $\g$
in different forms (see preceding Section V). We shall use
the form  of this tensor which is the most  close, in
construction, to the Minkowski energy-momentum tensor. In
the preceding Section this representation for the Abraham
tensor was denoted  by $T^{\,\text{A}2}_{\alpha\beta}$
(see Eq.\  \eqref{tA2}),
\begin{equation}
\label{ab-1}
T^{\,\text{A}2}_{\alpha\beta}=\tm +\frac{\varepsilon\mu-1}{4\pi}\,u_\beta\Omega_\alpha\,{,}
\end{equation}
where  $\tm$ is the Minkowski energy-momentum tensor
\eqref{tm-cov}, $\Omega_\alpha$ is the Minkowski vector
\eqref{R-S}, and $u_\alpha$ is the 4-velocity of the medium
\eqref{eu}. The right-hand-side of Eq.\ \eqref{ab-1} is not
symmetric with respect to indices $\alpha \beta$, however
it was shown  in  Sec.\ V that on $\g$ tensor
$T^{\,\text{A}2}_{\alpha\beta}$ coincides with  explicitly symmetric
tensor $T^{\,\text{A}1}_{\alpha\beta}$ \eqref{tA1} (see
Eq.\ \eqref{ttt}).

Differentiation of Eq.\ \eqref{ab-1} gives the
ponderomotive force  in the Abraham approach
\begin{equation}
\label{ab-2}
f^{\text{A}}_\alpha\equiv T^{\,\text{A}2}_{\alpha\beta,\beta}=f^{\,\text{M}}_\alpha+\frac{1}{4\pi }\,\{
(\varepsilon\mu-1)\Omega_\alpha u_\beta
\}_{,\beta}\,{,}
\end{equation}
where  $f^{\,\text{M}}_\alpha$ is the ponderomotive force
in the Minkowski approach \eqref{N-6}.

Thus it is required to calculate only  the difference
between ponderomotive forces in the Abraham and Minkowski
approaches
\begin{equation}
\label{ab-3}
\Delta f_\alpha\equiv f^{\,\text{A}}_\alpha - f^{\,\text{M}}_\alpha=\frac{1}{4\pi}\,\{
(\varepsilon \mu -1)\Omega_{\alpha} u_\beta
\}_{,\beta}\,{.}
\end{equation}

    In the rest frame $K'$ we have
\begin{align}
\Delta \mathbf{f}'&= \frac{1}{4 \pi c}\, \frac{\partial}{\partial t'}\,\{
(\varepsilon \mu -1)[\mathbf{E'H'}]
\}\,{,} \label{ab-4} \\
\Delta f'_4&= 0\,{.} \label{ab-4a}
\end{align}
Upon taking into account  the constitutive relations
\eqref{e0} in  Eq.\ \eqref{ga-2}, we see that Eqs.\
\eqref{ab-4} and \eqref{ga-2} coincide on $\g$.

The right-hand side  of Eq.\ \eqref{ab-3} is expressed in
terms of two 3-dimensional vectors: the medium velocity
$\mathbf{q=v/}c$ and the vector $\W$ introduced in~\eqref{N-3},
\begin{align}
4\pi \Delta \mathbf{f}&= (\mathbf{q}\bm{\nabla})\W+ \W \,\text{div}\,
\mathbf{q}+\frac{1}{c}\,\W_{,t},\label{ab-5} \\
\frac{4\pi}{\rmi}\, \Delta
f_4&=(\mathbf{q}\bm{\nabla})(\mathbf{q}\W)+(\mathbf{q}\W)\,\text{div}\,\mathbf{q}+\frac{1}{c}\,
(\mathbf{q}\W)_{,t}\,{.}  \label{ab-6}
\end{align}
Let us remind that  the vector  $\W$ is defined on  $\g$ by
Eq.\ \eqref{N-5} and does not depend explicitly  on the
medium characteristics $\varepsilon$ and $\mu$.

Equations \eqref{ab-3} and \eqref{ab-5}, \eqref{ab-6}
together with Eqs.\  \eqref{N-6}, \eqref{N-7}, and
\eqref{N-8} provide a complete solution of the problem
about the calculation of the ponderomotive force in
the Abraham approach. In order to  represent the covariant
formulae \eqref{ab-3} and \eqref{N-6} in terms of the same
variables it is sufficient  to substitute the vector
$N_\alpha$ in the form \eqref{N-2} into Eq.\ \eqref{N-6}.
It is worth noting that Eqs.\ \eqref{ab-2}--\eqref{ab-6}
are derived on the physical configuration space $\g$. It is
this circumstance that enables us to obtain compact and
convenient to  analyse formulae determining the
ponderomotive forces in the Minkowski and Abraham
approaches.

\section{Conclusion}
\label{Concl} In the general setting of the problem, the
explicit compact formulae are derived for the ponderomotive
forces in the Minkowski approach (Eqs.\
\eqref{N-6}--\eqref{N-8}) and in the Abraham approach
(Eqs.\ \eqref{ab-2}--\eqref{ab-6}). Two principal points in
calculation should be stressed: i) the formulae are derived
on the physical configuration space $\g$, i.e., with
allowance for the the Minkowski constitutive relations; ii)
a special representation for the Abraham energy-momentum
tensor \eqref{ab-1} is used which is the most close, in
construction, to the Minkowski tensor. In deriving the
ponderomotive forces the most general case is considered,
namely, nonhomogeneous nonstationary medium, the external
electromagnetic field and medium velocity can vary in space
and with time.

The formulae obtained (see Eqs.\
\eqref{ab-3}--\eqref{ab-6}) give a simple expression for
the difference of the ponderomotive forces in the
approaches under consideration. Thus we have derived the
generalization to the moving media of the known expression
for the Abraham force known for media at rest
\cite[\S~75]{LL8}.

An interesting result obtained in this paper is also the
derivation of the Lorentz force  which is exerted by
external electromagnetic field on the conduction current in
a medium. In deriving the formulae \eqref{N-11} and
\eqref{N-12} for this force the covariant Ohm law and the
constitutive Minkowski relations were taken into account.

\appendix
\section{Ponderomotive  forces in vacuum}
\label{A} The electric charges (with density $\rho$\,) and
currents (with density $\mathbf{j}$\,), placed in vacuum,
experience the Lorentz force
\begin{equation}
\label{a1}
f_\mu=\frac{1}{c}\,F_{\mu \nu}j_\nu
\end{equation}
from electromagnetic field $F_{\mu \nu}$ created by these
charges and currents. In Eq.\ \eqref{a1}
$j_\nu=(\mathbf{j},\rmi c\rho )$ and tensor $F_{\mu \nu}$
is obtained from  \eqref{m-7} by substituting $H_k$ in
place of $B_k$. All the charges and currents we treat on
the same footing without dividing them into those involved
in the system under consideration and external ones with
respect to this system. In addition, we digress from their
self-action \cite[sec.\ 42]{Pauli}, \cite[Chap.\
III]{Groot}.

Taking into account the Maxwell equations
in vacuum
\begin{gather}
F_{\alpha\beta,\gamma}+F_{\beta\gamma,\alpha}+F_{\gamma\alpha,\beta}=0\,{,} \label{a3}\\
F_{\alpha\beta,\beta}=\frac{4 \pi}{c}\,j_\alpha\,{,}
\label{a4}
\end{gather}
we can represent the Lorentz force \eqref{a1} as the
4-divergence of the energy-momentum tensor of
electromagnetic field in vacuum~\cite{Jackson,LL2}
\begin{equation}
f_\mu=T_{\mu\nu ,\, \nu}\,{,}
\label{a5}
\end{equation}
where
\begin{equation}
T_{\mu\nu}= \frac{1}{4\pi}\left (
F_{\mu\lambda}F_{\lambda\nu}-\frac{\delta_{\mu \nu}}{4}F_{\lambda\rho}F_{\rho\lambda}
\right )
{.}
\label{a6}
\end{equation}
Let us prove this assertion. Substitution of
\eqref{a6} into \eqref{a5} gives
\begin{equation}
4\pi f_\mu=
(F_{\mu\lambda}F_{\lambda \nu})_{,\nu}-\frac{1}{4}\,
(F_{\lambda\rho}F_{\rho\lambda})_{,\mu}\,{.}
 \label{a7}
\end{equation}
The first term on the right-hand side of \eqref{a7} we
transform by making use of Maxwell's equations \eqref{a3}
and \eqref{a4}
\begin{multline}
(F_{\mu\lambda}F_{\lambda\nu})_{,\nu} =F_{\mu\lambda}F_{\lambda \nu,\nu}+
F_{\mu\lambda,\nu}F_{\lambda\nu} \\
{}=\frac{4\pi}{c}\,F_{\mu\lambda}j_\lambda +\frac{1}{2}\, (F_{\mu\lambda,\nu}-
F_{\mu\nu,\lambda})
F_{\lambda\nu}
\\
{}=\frac{4\pi}{c}\,F_{\mu\lambda,}j_\lambda-\frac{1}{2}\,
F_{\lambda\nu,\mu}F_{\lambda\nu}=
\frac{4\pi}{c}\,F_{\mu\lambda,}j_\lambda+\frac{1}{4}\,
(F_{\lambda\nu}F_{\nu\lambda})_{,\mu}\,{.} \label{a8}
\end{multline}
Substitution of   \eqref{a8} in \eqref{a7} results in the
Lorentz formula \eqref{a1}
\begin{equation}
\label{a9}
f_\mu=T_{\mu\nu,\nu}=\frac{1}{c}\,F_{\mu\lambda}j_\lambda\,{.}
\end{equation}
These equalities can be interpreted in the following way. Two
expressions  for the Lorentz force  presented in \eqref{a9}
are  equivalent  with allowance for  Maxwell's equations
\eqref{a3}, \eqref{a4}. Let us write down the components of
these expressions for the Lorentz force. To this end it is
convenient to use the standard notations for the individual
components of the energy-momentum tensor  $T_{\mu \nu}$ \eqref{a6}:
\begin{equation}
\label{a10} T_{\mu\nu}=\left (
\begin{matrix}
\sigma_{ij}&-\rmi c\,
\mathbf{g}\cr -\frac{\displaystyle
\rmi}{\displaystyle  c}
\,\mathbf{S}^{\text{P}}&w \cr
\end{matrix}
\right ){,}
\end{equation}
where
\begin{gather}
\sigma_{ij}=\frac{1}{4 \pi}\left \{
E_i E_j+H_i H_j-\frac{\delta_{ij}}{2}\,(\mathbf{EE+HH})
\right \}{,}\label{a11} \\
\frac{1}{c}\,
 \mathbf
{S}^{\text{P}}=\frac{c}{4\pi}\,[\mathbf{EH}]\,{,}\quad
c\mathbf {g}= \frac{1}{c}\,\mathbf{S}^{\text{P}}\,{,}
\label{a12} \\
w=\frac{1}{8
\pi}\,(\mathbf{EE+HH})\,{.}\label{a13}
\end{gather}
Here $\mathbf{S}^{\text{P}}$ is the Poynting vector \eqref{P}.

On substituting \eqref{a10} in \eqref{a9}, we get
\begin{align}
\mathbf{f}&=\bm{\sigma}-\mathbf{\dot g}=\rho
\mathbf{E}+\frac{1}{c}\, [\mathbf{jH}]\,{,}
\label{a14}\\
f_4&=-\frac{\rmi}{c}\,(\text{div}\,\mathbf{S}^{\text{P}}+\dot
w)=\frac{\rmi}{c}\,(\mathbf{jE})\,{,} \label{a15}
\end{align}
where $\bm{\sigma}\equiv (\sigma_1,\sigma_2,\sigma_3),\quad
\sigma_i=\sigma_{ik,k}$. For brevity the subscript
vac is omitted in this Appendix.

The foregoing reasoning devoted to the derivation of Eq.\
\eqref{a9}, proceeding from the given form of the
energy-momentum tensor \eqref{a6}, can be inverted, namely,
one can find the energy-momentum tensor \eqref{a6},
requiring the fulfilment of the second equality in Eq.\
\eqref{a9} with allowance of the Maxwell equations. Let us
prove this assertion. We replace in \eqref{a9} the current
$j_\lambda$ by making use of nonhomogeneous Maxwell's
equations \eqref{a4}:
\begin{align}
T_{\mu\nu,\nu}&=\frac{1}{c}\, F_{\mu\lambda}\frac{c}{4\pi}
\,F_{\lambda\rho,\rho}\nonumber \\
&=\frac{1}{4\pi}\,(F_{\mu\lambda}F_{\lambda\rho})_{,\rho}-\frac{1}{4\pi}
\,F_{\mu\lambda,\rho}F_{\lambda\rho}\,{.}
\label{a16}
\end{align}
Further with the help of homogeneous Maxwell's equations
 \eqref{a3} we can write
\begin{equation}
-F_{\mu\lambda,\rho}F_{\lambda\rho}=-\frac{1}{2}\,(F_{\mu\lambda,\rho}
+F_{\rho \mu,\lambda})F_{\lambda\rho}=\frac{1}{2}\,
F_{\lambda\rho,\mu}F_{\lambda\rho}\,{.}
\label{a17}
\end{equation}
From \eqref{a14} and  \eqref{a15} it follows
\begin{equation}
4\pi T_{\mu\nu,\nu}=(F_{\mu \lambda}F_{\lambda\nu})_{,\nu}-\frac{1}{4}\,
(F_{\lambda\rho}F_{\rho\lambda})_{,\mu}\,{.}
\label{a18}
\end{equation}
Obviously  tensor $T_{\mu \nu}$  in the form \eqref{a6}
satisfies Eq.\ \eqref{a18}.

Thus one can conclude  that the expression for the Lorentz
force \eqref{a1} and the explicit form  of the
energy-momentum tensor of the electromagnetic field in
vacuum \eqref{a6} (the Lorentz theory of electrons) are
equivalent in the meaning specified above.

In the macroscopic electrodynamics (in the theory of
electromagnetic field in a medium) the situation is
essentially different.  As was noted in Section \ref{ftei},
it does  not turn out well to obtain the expression  for
the ponderomotive forces exerting on the medium (the
analogue of Eq.\ \eqref{a1}) proceeding from the Lorentz
theory of electrons. The analogy with the electromagnetic
field in vacuum can provide only  the definition of the
ponderomotive forces through the energy-momentum tensor
\eqref{a5}. This ensures  the covariance of the  definition
and reduces the problem to looking for the macroscopic
energy-momentum tensor of electromagnetic field.
\section{The explicit formula for vector $\W$}
\label{B} In order to find the explicit expression for the
vector  $\W$ on $\g$ (see Eq.\ \eqref{N-3}) we take
advantage of the equality valid on $\g$ \cite[p.\ 51,
Eq.\ (119)]{Kafka}:
\begin{equation}
\label{w-4} (\varepsilon
\mu-1)(u_\alpha\Omega_{\beta}-u_\beta\Omega_{\alpha})\left
|_{\overline \Gamma}=F_{\alpha\gamma} H_{\gamma
\beta}-F_{\beta\gamma }H_{\gamma \alpha}\right |_{\overline
\Gamma}\,{.}
\end{equation}
First we prove this equality by the method different from
that used in Ref.\ \cite{Kafka}, namely, we demonstrate
that on $\g$ the right-hand side of \eqref{w-4} transforms
into the left-hand side of this formula.  To this end we
employ the representations \eqref{Kf-1rr} and
\eqref{Kf-2rr} for the tensors $F_{\alpha \beta}$ and
$H_{\alpha \beta}$ which are valid on $\g$:
\begin{equation}
\label{w-5}
F_{\alpha \beta}= {\cal E}_{\alpha \beta }+\rmi \mu {\cal H}_{\alpha \beta }\,{,}
\quad
H_{\alpha \beta}=\varepsilon {\cal E}_{\alpha \beta }+
\rmi  {\cal H}_{\alpha \beta}\,{,}
\end{equation}
where ${\cal E}_{\mu \nu}$ and ${\cal H}_{\mu \nu}$ are
determined  in \eqref{EH}. Substitution of  \eqref{w-5}
into the right-hand side of \eqref{w-4} yields
\begin{gather}
\left . F_{\alpha\gamma} H_{\gamma \beta}-F_{\beta\gamma }H_{\gamma
\alpha}\right |_{\overline \Gamma}=({\cal
E}_{\alpha\gamma}+\rmi \mu {\cal
H}_{\alpha\gamma})(\varepsilon {\cal E}_{\gamma\beta}+\rmi {\cal H}_{\gamma\beta})
 -(\alpha \leftrightarrow \beta) \nonumber \\
= \varepsilon {\cal E}_{\alpha\gamma}{\cal
E}_{\gamma\beta}-\mu {\cal H}_{\alpha\gamma}{\cal
H}_{\gamma\beta}+ \rmi \varepsilon \mu {\cal
H}_{\alpha\gamma}{\cal E}_{\gamma\beta} +\rmi{\cal
E}_{\alpha\gamma}{\cal H}_{\gamma\beta}-(\alpha
\leftrightarrow \beta)\,{.}\label{w-6}
\end{gather}
Two first terms in the right-hand side of \eqref{w-6},
which are symmetric  with respect to indexes $\alpha,
\beta$, are canceled with the analogous terms in the
expression denoted by $(\alpha
\leftrightarrow \beta)$. Therefore it needs to
calculate explicitly only the product ${\cal
E}_{\alpha\gamma}{\cal H}_{\gamma\beta}$. With the help of
the definitions \eqref{EH}, we obtain
\begin{align}
{\cal E}_{\alpha\gamma}{\cal H}_{\gamma\beta}&=(u_\alpha
E_\gamma -u_\gamma
E_\alpha)\varepsilon_{\gamma\beta\delta\rho}u_\delta H_\rho
\nonumber \\
&=u_\alpha \varepsilon_{\beta\gamma\rho\delta}E_\gamma
H_\rho u_\delta =\rmi u_\alpha \Omega_\beta \,{,}\label{w-7}
\end{align}
where  $\Omega_\beta$ is the Minkowski vector \eqref{R-S}. Substitution of
\eqref{w-7} in \eqref{w-6} gives
\begin{align}
\left .F_{\alpha\gamma}
H_{\gamma\beta}-F_{\beta\gamma}H_{\gamma\alpha}\right
|_{\overline \Gamma}&= -u_\alpha \Omega_\beta+u_\beta \Omega_\alpha
 -\varepsilon\mu u_\beta \Omega_\alpha +
\varepsilon \mu u_\alpha \Omega_\beta \nonumber \\
&=(\varepsilon \mu -1)\left .(u_\alpha\Omega_\beta -u_\beta \Omega_\alpha) \right |
_{{\overline \Gamma}}\,{.}
\label{w-8}
\end{align}
Thus the equality \eqref{w-4} is proved.

The components of the antisymmetric tensors in the
left-hand side and in the right-hand side  of Eq.\
\eqref{w-4} are determined by two 3-dimensional vectors
(see  notations introduced in \eqref{m-7}--\eqref{m-8a}).
Therefore  Eq.\ \eqref{w-4} is equivalent to two
3-dimensional vector equalities. We write down these
equalities in an explicit form by making use of Eq.\
\eqref{eu} and definition  \eqref{N-3}.

When the pair of indexes $\alpha\,\beta$ in Eq.\
\eqref{w-4} takes the values $1\,4,\;2\,4,\;3\,4$ the
following vector equality arises here:
\begin{equation}
\label{w-9}
-\rmi\{\W-\mathbf{q}(\mathbf{q}\W)\}=-\rmi\{\mathbf{[DB]-[EH]}\}
\end{equation}
or in another form
\begin{equation}
\label{w-10}
\W=\mathbf{[DB]-[EH]}+\mathbf{q}(\mathbf{q}\W)\,{.}
\end{equation}
From Eq.\
\eqref{w-10} it folows
\begin{eqnarray}
(\mathbf{q}\W)&=&(\mathbf{q,[DB]-[EH]})+q^2(\mathbf{q}\W)= \nonumber \\
&=& \gamma^2(\mathbf{q,[DB]-[EH]})\,{.}\label{w-11}
\end{eqnarray}
Finally we arrive at the result
\begin{equation}
\label{w-12}
\W=\mathbf{[DB]-[EH]}+\gamma^2 \mathbf{q (q, [DB]-[EH])}\,{.}
\end{equation}

If in Eq.\  \eqref{w-4}
the pair of indexes  $\alpha\,\beta$ assumes the values
$2\,3,\; 3\,1, \;  1\,2$, then we obtain another vector equality
\begin{equation}
\label{w-13}
[\mathbf{q}\W]=\mathbf{[ED]+[HB]}\,{.}
\end{equation}
Substitution of  \eqref{w-12} into  \eqref{w-13} yields
\begin{equation}
\label{w-14}
\mathbf{[q,[DB]-[EH]]=[ED]+[HB]}\,{.}
\end{equation}

We stress once more that the obtained equations
 \eqref{w-12}--\eqref{w-14}, as well as the initial
equality \eqref{w-4}, are valid only under fulfilment of
the Minkowski relations, i.e., only on $\g$.


\begin{thebibliography}{99}
\bibitem{Brevik-1} I.\ Brevik, Mat.\ Phys.\ Medd.\ Dan.\ Vid.\ Selsk.\
{\bf 37}(11), 1 (1970); {\bf 37}(13), 1 (1970).
\bibitem{Brevik-3} I.\   Brevik,  Phys.\ Rep.\ {\bf 52}, 133 (1979).
\bibitem{NN-A}  V.\ V.\ Nesterenko and A.\ V.\ Nesterenko,
J.\  Math.\  Phys.\ {\bf 57}, 032901 (2016).
\bibitem{Nelson} D.\ F.\ Nelson, Phys.\ Rev.\ A {\bf 44}, 3985 (1991).
\bibitem{Groot} S.R.~de~Groot and L.G.~Suttorp,
{\it Foundations of Electrodynamics}
(North-Holland, Amstredam, 1972), Chap.~V, \S~7.
\bibitem{Obukhov} Yu.\ N.~Obukhov, Ann.\ Phys.\ (Berlin), {\bf 17}, No.~9/10,
830 (2008).
\bibitem{MR} V.\ P.~Makarov and A.\ A.~Rukhadze, Phys.\ Usp.\ {\bf 52}, 937 (2011)
[Usp.\ Fiz.\ Nauk {\bf 181}, 1357 (2011)].
\bibitem{Pauli}  W.\ Pauli, {\it Relativit\"atstheorie} (Teubner, Leipzig,  1921)
[W.\ Pauli, {\it Theory of Relativity} (Pergamon Press, New York, 1958)].
\bibitem{M-zwei} H.\ Minkowski,
{\it Zwei Abhandlungen \"uber die Grundgleichungen der Elektrodynamik}
(B.G.\ Teubner, Leipzig und Berlin, 1910).
\bibitem{Abraham1920}
M.\  Abraham,  {\it Theorie der Elektrizit\"at}, Bd.\ 2
(4.\ Aufl., Teubner, Leipzig, 1920).
\bibitem{AP1909} M.\  Abraham,  Palermo Rend.\ {\bf 28}, 1 (1909)
[http://en.wikisource.org/wiki/Author:Max\underline{~}Abraham].
\bibitem{AP1910} M.\  Abraham, Palermo Rend.\ {\bf 30}, 33 (1910)
[http://en.wikisource.org/wiki/Author:Max\underline{~}Abraham].
\bibitem{LL8}
L.\ D.\ Landau, E.\ M.\ Lifshitz, {\it Electrodynamics of Continuous Media}
 (Pergamon Press, Oxford, 1984).
\bibitem{LL2} L.\ D.\ Landau, E.\ M. Lifshitz, {\it The Classical Theory of Fields}
 (Pergamon Press, Oxford, 1980).
\bibitem{Kafka} H.\ Kafka,  Ann.\ Physik {\bf 58}, 1 (1919).
\bibitem{Laue} Max von Laue, {\it Das Relativit\"atstheorie}.
 Erster Band. {\it Das Relativit\"atsprinzip der Lorentz-transformation}.
 Vierte vermehrte Auflage
(Friedr.\ Vieweg \& Sohn, Braunschweig, 1921).
\bibitem{Becker}
R.\ Becker, {\it Theorie der Elektrizit\"at.} Band 2: {\it Elektronentheorie}
 (Teubner, Berlin-Leipzig, 1933).
\bibitem{Moller} Chap.\ M\o ller, {\it The theory of relativity}
(Clarendon Press, Oxford, 1972).
\bibitem{Herg}G.\ Herglotz, Ann.\  Physik {\bf 36},  497 (1911).
\bibitem{NYa} Yu.\ V.\ Novozhilov, Yu.\ A.\ Yappa, {\it Electrodynamics}
(Mir Publishers, Moscow, Russia, 1986).
\bibitem{Stratton} J.\ A.\ Stratton, {\it Electromagnetic Theory}
(Adams Press, Chicago, Illinois, USA, 2008).
\bibitem{PPh} W.\ K.\ H.\ Panofsky, M.\ Phillips,
{\it Classical Electricity and Magnetism}, 2nd ed.\ (Dover
Books on Physics, 2005).
\bibitem{Tamm} I.\ E.\ Tamm, {\it Fundamentals of the
Theory of Electricity} (Mir Publishers, Moscow, Russia,
1979).
\bibitem{Jackson} J.\ D.\ Jackson,
{\it Classical Electrodynamics}, 3rd edn (Wiley, New
York, 1998).
\bibitem{Schmutzer} E.~Schmutzer, {\it Relativistische Physik}
(Akademie Verlag, Leipzig, 1968) p.\ 408.
\bibitem{Grammel} R.\ Grammel,  Ann.\ Physik {\bf 346}, 570 (1913).
\end{thebibliography}
\end{document}